\documentclass[twocolumn,a4paper]{article}
\usepackage{nolta2012}
\usepackage{txfonts}
\usepackage{graphicx}

\begin{document}
\bibliographystyle{plain}

\title{A methodology for detecting and exploring non-convulsive
  seizures in patients with SAH}

\author{${DJ Albers}^\dag$ and ${J Claassen}^\ddag$and ${M
    Schmidt}^\ddag$ and ${G Hripcsak}^\dag$}

\address{
\dag Department of Biomedical Informatics, Columbia University\\
622 West 168th Street VC-5, New York, NY, 10032, USA\\
\ddag Department of Neurology, Columbia University\\
710 West 168th Street, New York, NY, 10032, USA\\
Email: david.albers@dbmi.columbia.edu, jc1439@mail.cumc.columbia.edu,
mjs2134@mail.cumc.columbia.edu, \\ hripcsak@columbia.edu
}

\maketitle

\abstract A methodology for understanding and detecting nonconvulsive
seizures in individuals with subarachnoid hemorrhage is
introduced. Specifically, beginning with an EEG signal, the power
spectrum is estimated yielding a multivariate time series which is
then analyzed using empirical orthogonal functional analysis. This
methodology allows for easy identification and observation of seizures
that are otherwise only identifiable though expert analysis of the raw
EEG.  \endabstract

\section{Introduction}

Seizure detection in EEG data has a long history
\cite{santa_fe_time_series_prediction,larry_neuroscience_book,partha_brain_book}. In
fact, seizure detection is developed to the point where medical
instrument companies have proprietary seizure detection
algorithms. Nevertheless, automated seizure detection is not
particularly effective. Moreover, most seizure detection is carried
out in the context of epilepsy. Here we develop a new methodology for
understanding and detecting seizure in a different context, in
patients with a aneurysmal subarachnoid hemorrhage (SAH). The context
is important because SAH is a serious clinical condition that affects
a broad population. And the context is different from the more
standard epilepsy context; seizures in individuals with SAH are not
well understood, may be diverse in type, and may affect recovery in
different ways. Because of the potential diversity in seizure in SAH
patients, we aim to both detect seizure events and understand and
phenotype the different seizure types. Our methodology is based
applying several targeted levels of analysis to the original EEG
signal; here this methodology entails estimating a power spectrum from
EEG data and then applying empirical orthogonal functional analysis
(EOF)to the power spectrum (PS).

\section{Seizures in individuals with SAH}

Aneurysmal subarachnoid hemorrhage (SAH) occurs when blood enters the
subarachnoid space, located between the arachnoid membrane and the pia
mater surrounding the brain, from a ruptured dilated cerebral blood
vessel. SAH affects up to 30,000 Americans annually carrying a huge
public health burden. Secondary complications such as nonconvulsive
seizures (NCSz) contribute significantly to poor outcome. These
seizures are different from convulsive seizures as the patients have
no or minimal symptoms other than decreased mental status while the
brain is seizing. There is a great deal of evidence that suggests that
additional brain injury occurs secondary to NCSz
\cite{subarachnoid_seizures_I}.

Treatment is available but diagnosis poses major challenges as
automated detection algorithms to date have very poor
accuracy. Controversy exists regarding the preferred treatment regimen
but unanimously experts agree that the time to initiate treatment is
much more important than the choice of seizure medication. Detection
algorithms fall short as surface EEG is notoriously contaminated by
artifact. Some sources of artifacts include poor contact between EEG
electrodes and scalp, sweat artifact, electrical artifact, and many
more. Intracrotical depth electrodes are increasingly placed together
with other invasive brain monitoring devices and have the huge
advantage of a better signal to noise ratio. Signals from such sources
would be an ideal to further develop seizure detection algorithms with
better specificity and sensitivity for seizures.

\paragraph{Diversity among seizures}

Seizures after acute brain injury show a great deal of phenotypical
variability. For example, approximately half of the seizures remain
focal and do not spread to other brain regions. Patterns of seizures
further differ greatly in terms of maximum frequency, amplitude,
duration, and background in between discharges. The pathophysiological
significance of these differences is unknown but to study these
differences accurate characterization of patterns is the first step.


\section{Seizure detection and analysis}

\begin{figure}
\includegraphics[width=7cm,height=3cm]{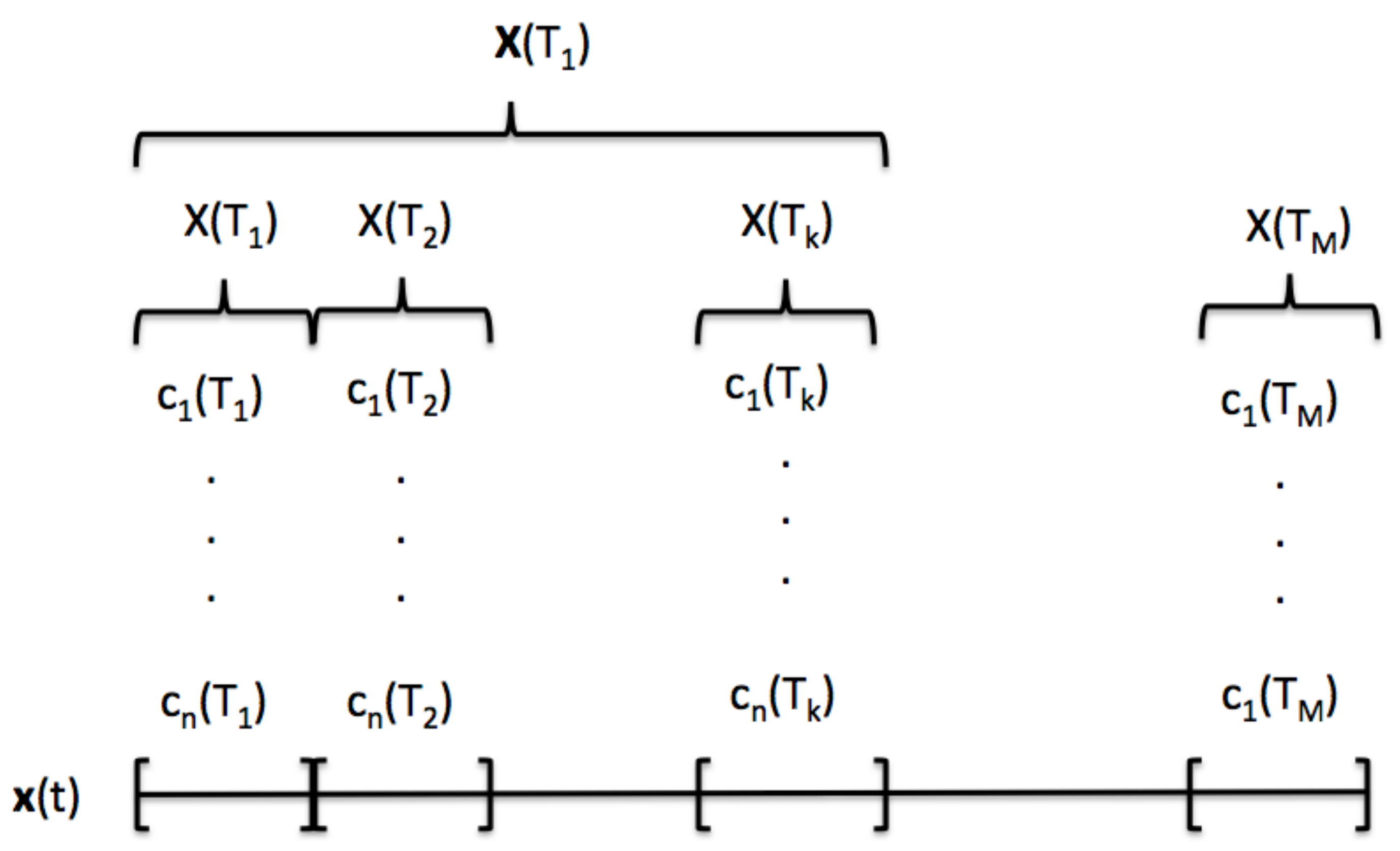}
\label{fig:broad_schematic}
\caption{\footnotesize Broad view of our seizure analysis methodology. Given a time
  series ($\mathbf{x}(t)$), we estimate a power spectra at fixed
  intervals yielding powers per frequency ($c_i(T_j)$) that are then
  treated as a multi-variate time series ($\mathbf{X}(T_i)$) that is
  then studied using EOF analysis.}
\end{figure}



\paragraph{General decomposition of a time series}
Begin with a time series of length $MT$, $\mathbf{x}(MT)=(x(1), x(2),
\cdots, x(MT))$. Assume over the time period or window of calculation
that the time series is ergodic and obeys the weak stationarity
property \cite{koopmans,statistical_analysis_climate_book}. We will,
for now, split the time series $\mathbf{x}(MT)$ into $M$ components,
indexed by $k$ and denoted by $\mathbf{x}_k(t)$. The time series can
be decomposed into $n$ orthogonal components, or
\begin{equation}
\label{equation:general_decomp}
\mathbf{x}_k(t) = \sum_{i=j}^N c_j \mathbf{e}_j
\end{equation}
$\mathbf{e}_j$ is an orthogonal basis functions and $c_j$ is the
amplitude of that given basis function. In this paper, we will begin
with a time series of finite length, $MT$, split it into $M$
components of length $T$, and then decompose the disjoint time series
in two different steps to achieve a useful representation of seizure
dynamics.

\paragraph{Power spectra decomposition of EEG data}

It is well known \cite{koopmans} that a way to represent a time series
is via a decomposition into a set of $2n$ frequencies, $\lambda_j$, or
\begin{equation}
\mathbf{x}_k(t)= \sum_{j=-n}^{n} c_j e^{i \lambda_j t};
\end{equation}
this representation of $\mathbf{x}_k(t)$ is called the Fourier
transform \cite{whezyg,koopmans,fourierO_dirichlet} of
$\mathbf{x}_k(t)$. In this framework $\mathbf{x}_k(t)$ is
conceptualized as a collection of harmonic terms parameterized by
frequency. Each frequency has an instantaneous power associated with
it $|c_j|^2$; or $|c_j|^2$ is the power of $x(t)$ at frequency
$\lambda_j$. Intuitively the power at a frequency quantifies how much
of $\mathbf{x}(t)$'s signal is represented by the orthogonal
component, or harmonic term, $c_j e^{i \lambda_j t}$, over the time
window of length $T$. The power calculation yields the vector of
powers for given frequencies, $X_k(t) = (|c_1|^2, |c_2|^2, \cdots,
|c_n|^2)$, for each of the $M$ time series. Finally, by Parseval's
theorem \cite{whezyg,koopmans} the total power in frequency space is
equal to the variance in state space, or:
\begin{eqnarray}
\sigma(\mathbf{x}_k(T)) = P(\mathbf{x}_k(T)) &= \frac{1}{T}
\sum_{t=1}^T x^2(t) \\
  &= \sum_{j=-n}^{n} |c_j|^2
\end{eqnarray}


\paragraph{Empirical orthogonal functional analysis}

To study the time series of the vector of power per frequency,
$\mathbf{X}(T_k)$, a multivariate time series, we must generalize to
multivariate, or matrix decompositions. For the moment we will ignore
the meaning of the time points (powers) and abstract the $|c_j|^2$'s
to be any variable dependent on time.

Consider a matrix of time series, where the columns index the time
points, and the rows index the variables (e.g., the
frequencies). Assume we have de-trended $\mathbf{X}(T_k)$ such that it
is mean zero. Associated with $\mathbf{X}(T_k)$ is its covariance
matrix, $\Sigma(T_k)$. By construction $\Sigma$ is Hermitian so the
eigenvalues are non-negative and there will always exist an orthogonal
basis (i.e., the eigenvectors are orthogonal). We can decompose
$\mathbf{X}(T_k)$ by the orthogonal directions of maximum variance
which is equivalent to (cf. section 13.1 in
\cite{statistical_analysis_climate_book}) decomposing
$\mathbf{X}(T_k)$ into eigenvectors (e.g., the generalized
$\mathbf{e}_j$'s from Eq. \ref{equation:general_decomp}) corresponding
to the eigenvectors of $\Sigma(T_k)$ in descending order. In this way,
the first EOF is the pattern representing the maximum variance within
$\mathbf{X}(T_k)$; written more mathematically, the first EOF is the
eigenvector that minimizes $E(||\mathbf{X}-\langle
\mathbf{X},\mathbf{e} \rangle \mathbf{e}||^2)$. Note that
$\mathbf{X}(T_k)$ must be at least a full rank matrix which in
practice means that $k\geq n$.

We visualize the EOFs individually as vectors, usually restricted to
the first EOF, creating a multivariate time series of length
$M$. Moreover, to we also estimate and plot the time-dependent
fraction of energy or variance represented by the EOF being
plotted. Recall that the total power (variance), is the sum of the
eigenvalues (the trace) of $\Sigma(T_k)$, or
$\sigma(\mathbf{X}(T_k))=\sum_{j=1}^n \lambda_j$. Using this, the
fraction of the variance that EOF $j$ represents is
$\frac{\lambda_j}{\sigma(\mathbf{X}(T_k)}$.


\paragraph{Explicit details of the PS and EOF computations}

\begin{figure*}
\centering
\includegraphics[width=5cm,height=3cm]{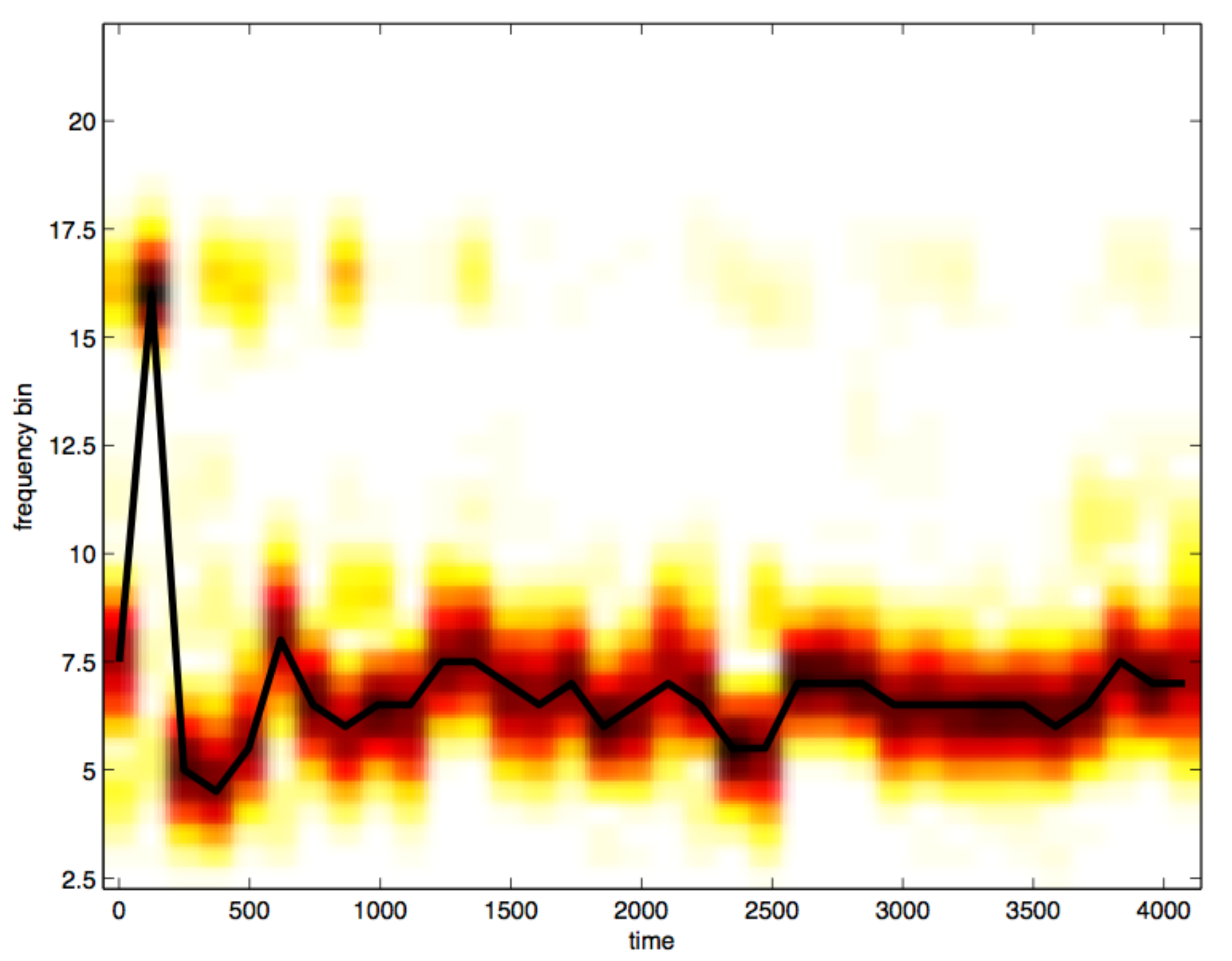}
\includegraphics[width=5cm,height=3cm]{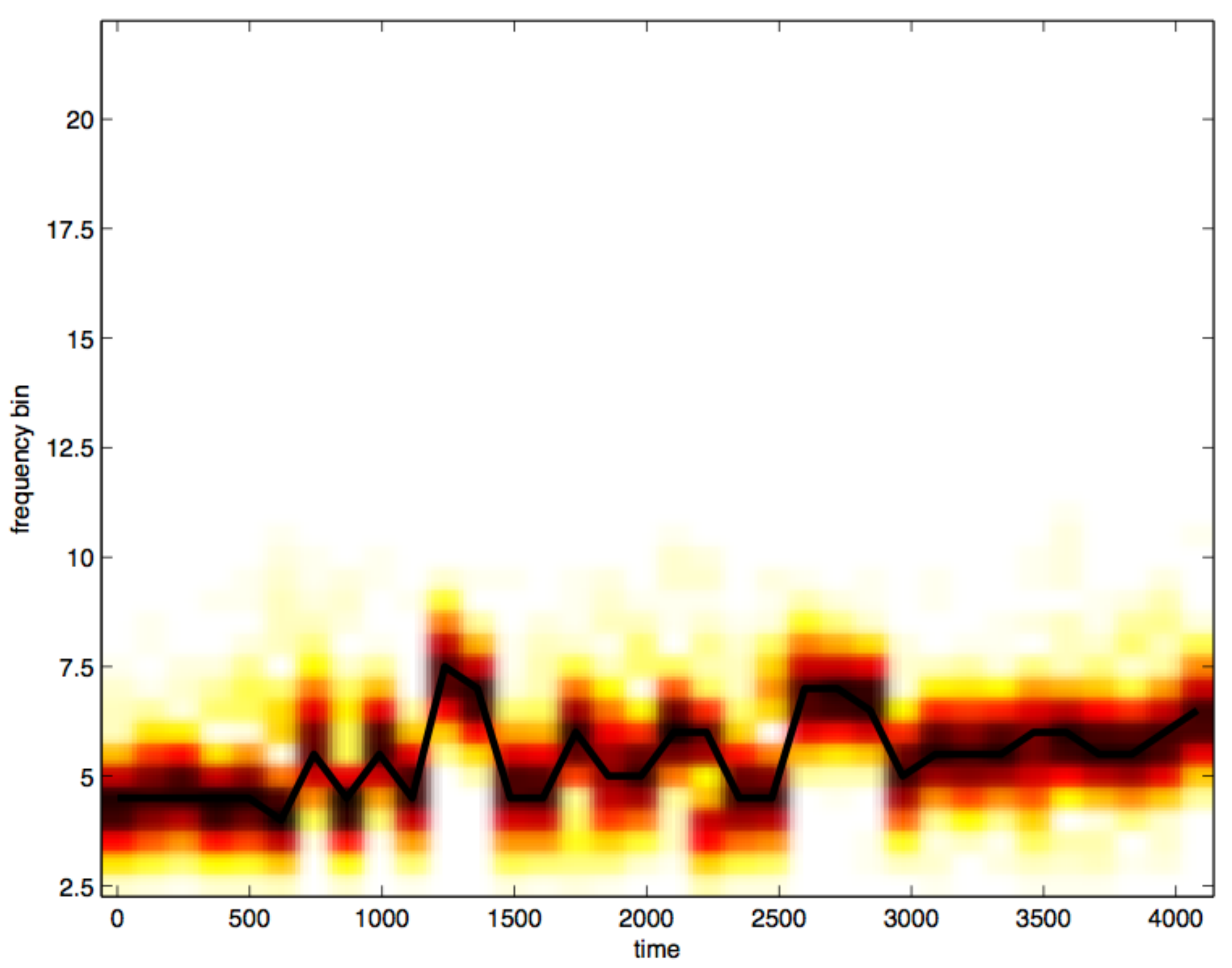}
\includegraphics[width=5cm,height=3cm]{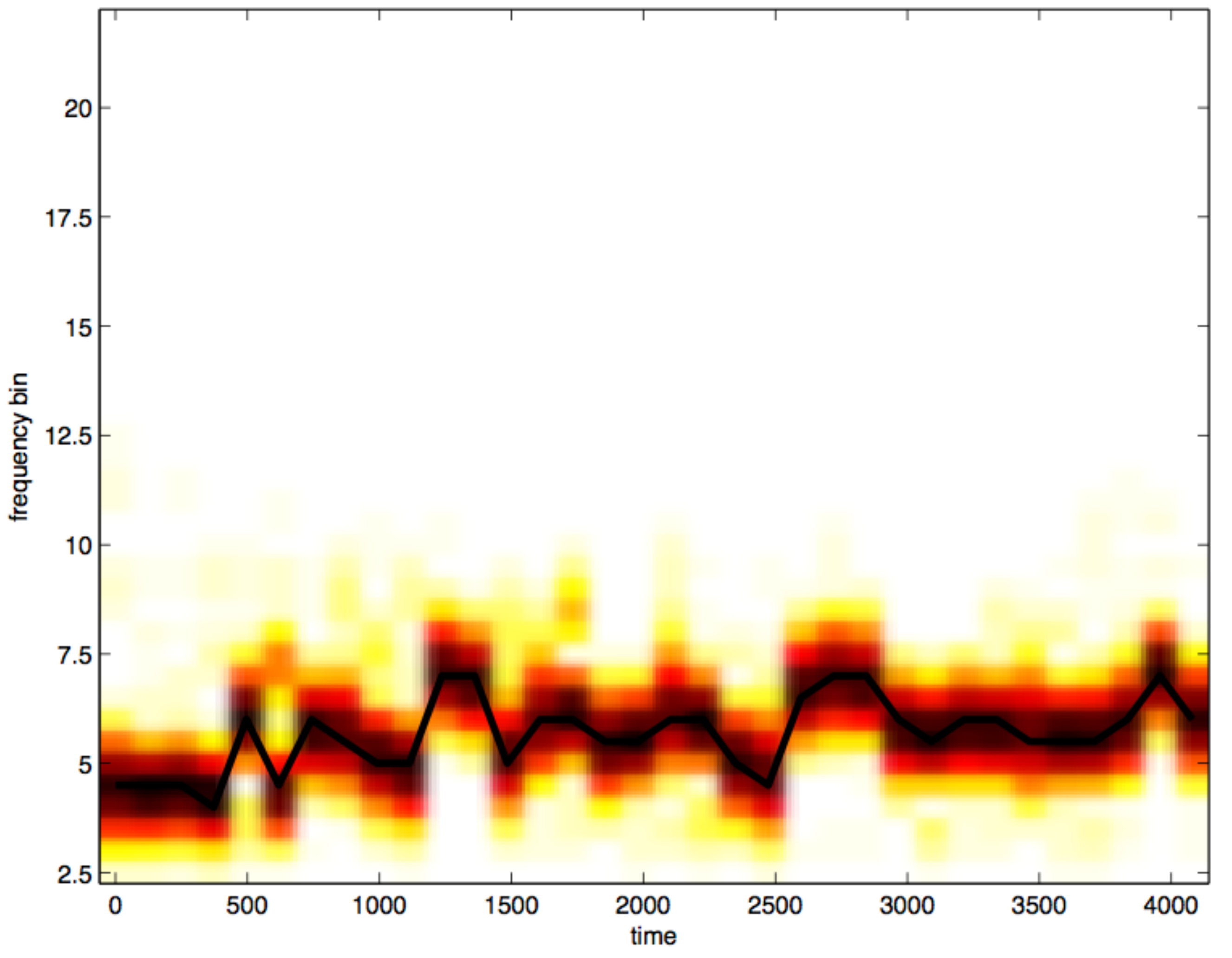}
\caption{\footnotesize EOF of the PS of the EEG for the left, right, and
  whole brain (left to right) respectively; note depth seizure event
  is only apparent in the EOF of the \emph{left} side of the brain and
  is indicated by a qualitative change in the EOF
  signal as realized via a larger number and different set of
  frequencies contributing to the variance of the signal.
}
\label{fig:eof_weak}
\vspace{-0.5cm}
\end{figure*}

The explicit algorithm we used in this paper follows four steps:
\emph{(i)} collect a time series of PS data, which is estimated by the
machinery used to collect EEG data; \emph{(ii)} determine a suitable
EOF window size, noting that $k\geq n$, thus creating the
$\mathbf{X}(T_j)$'s; \emph{(iii)} estimate the EOF on
\emph{non-overlapping} time windows of the PS data, the
$\mathbf{X}(T_j)$'s; and \emph{(iv)} plot the first EOF, a
$n$-dimensional vector, in time (cf. Figs. \ref {fig:EOF} and
\ref{fig:eof_weak}).

\paragraph{Interpretation of the EOFs of the time series of PS of a time series}

Moving back into the context where $\mathbf{X}(T_k)$ is a matrix of
time series of power per frequency, the interpretation of PS, the
first EOF, and the first EOF of the PS are as follows.

\emph{Power spectrum of the time series:} a time series can be
decomposed and represented as energy or power (if integrated over
time) of the frequencies that compose the time series.  Or, written
differently, the time series can be rewritten in terms of power
(variance) per orthogonal basis element.  In this situation, the basis
elements are \emph{not chosen to maximize any quantity}, but rather as
the frequencies present in the data.

\emph{First EOF of a multivariate time series:} the orthogonal
direction that represents the direction of greatest (or maximized)
energy or variance and can include portions of different frequencies
(e.g., $3$ Hz accounts for $20 \%$ , $5$ Hz accounts for $40 \%$,
etc.); the first EOF is a vector that specifies what proportions of
which frequencies make up the orthogonal vector in the direction of
the maximum energy or variance.

\emph{First EOF of the PS of a time series:} the ranked (or ordered)
proportional selection of frequencies that contribute the most energy
without overlapping (i.e., along the orthogonal component that
captures or represents the maximum amount of energy).  This is
equivalent to identifying the frequencies, by proportion, that
contribute the most to variance, or energy in the EEG signal. Because
variance and energy are synonymous, the first EOF of the PS of a time
series reveals the frequencies within the EGG signal that are the most
active, important, present, or energetic. 



\section{Results}

\paragraph{Identifying high resolution EOF features before and after
  manually identified seizure events.}

\begin{figure}
\centering
\includegraphics[width=7cm,height=4cm]{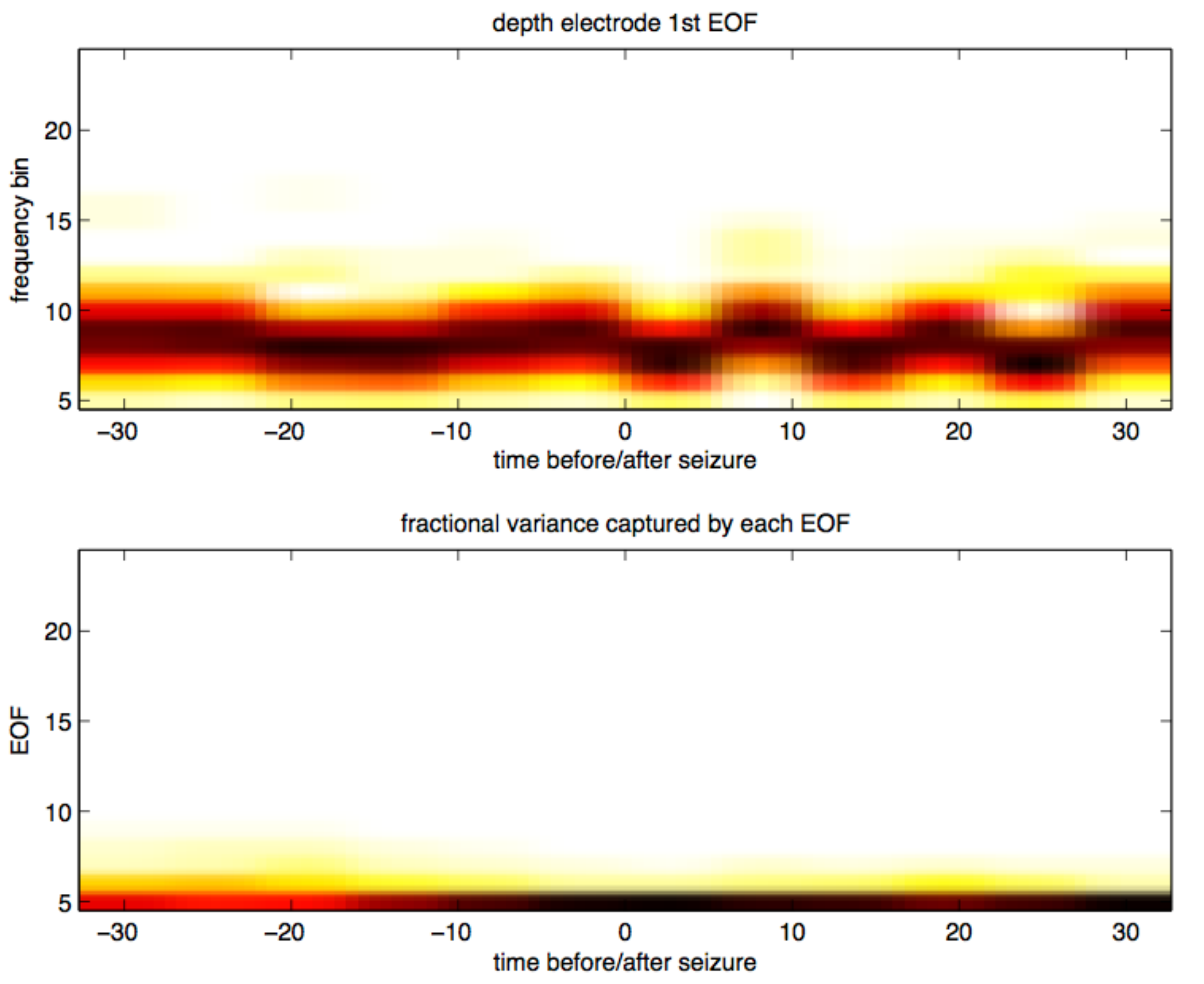}
\caption{\footnotesize First EOF (top) and relative importance (fractional
  variance) of each EOF versus time before and after seizure for power
  spectra of an EEG time series. Blue
  implies no contribution, red implies maximum contribution to the EOF.}
\label{fig:EOF}
\end{figure}

To study the nature of depth seizures in SAH patients, we applied EOF
analysis to the PS of the EEG for $n=20$ ranging from $1$ to $20$ Hz
recorded at $5$ second intervals ($T=5 min$, $k=60$) $30$ minutes
before and after seizure ($M=60 min$) for several single patient with
a SAH; because $k=60$ and $n=20$, without pathologies
$\mathbf{X}(T_k)$ will be full rank. The seizure onsets were manually
identified by a neurointensivist. The signal via EOF of the PS of the
EEG revealed is striking: at the onset of seizure, the period of
oscillation between the high-variance frequencies changes; in the
specific patient shown in Fig. \ref{fig:EOF} from the oscillation in
frequency changes period from $25$ to $12$ minutes. Moreover, as the
seizure onsets, the amount of variance represented by the first EOF
increases dramatically, as can be seen in Fig. \ref{fig:EOF}. This
implies that as the during the seizure, the set of excited frequencies
is severely constrained. It is hoped that temporal signatures such as
these can be generalized to better understand seizure, and to
phenotype different types of seizure in SAH patients.

\paragraph{Seizure visualization and detection prior to manual seizure
identification.}

\begin{figure*}
\centering
\includegraphics[width=5cm,height=3cm]{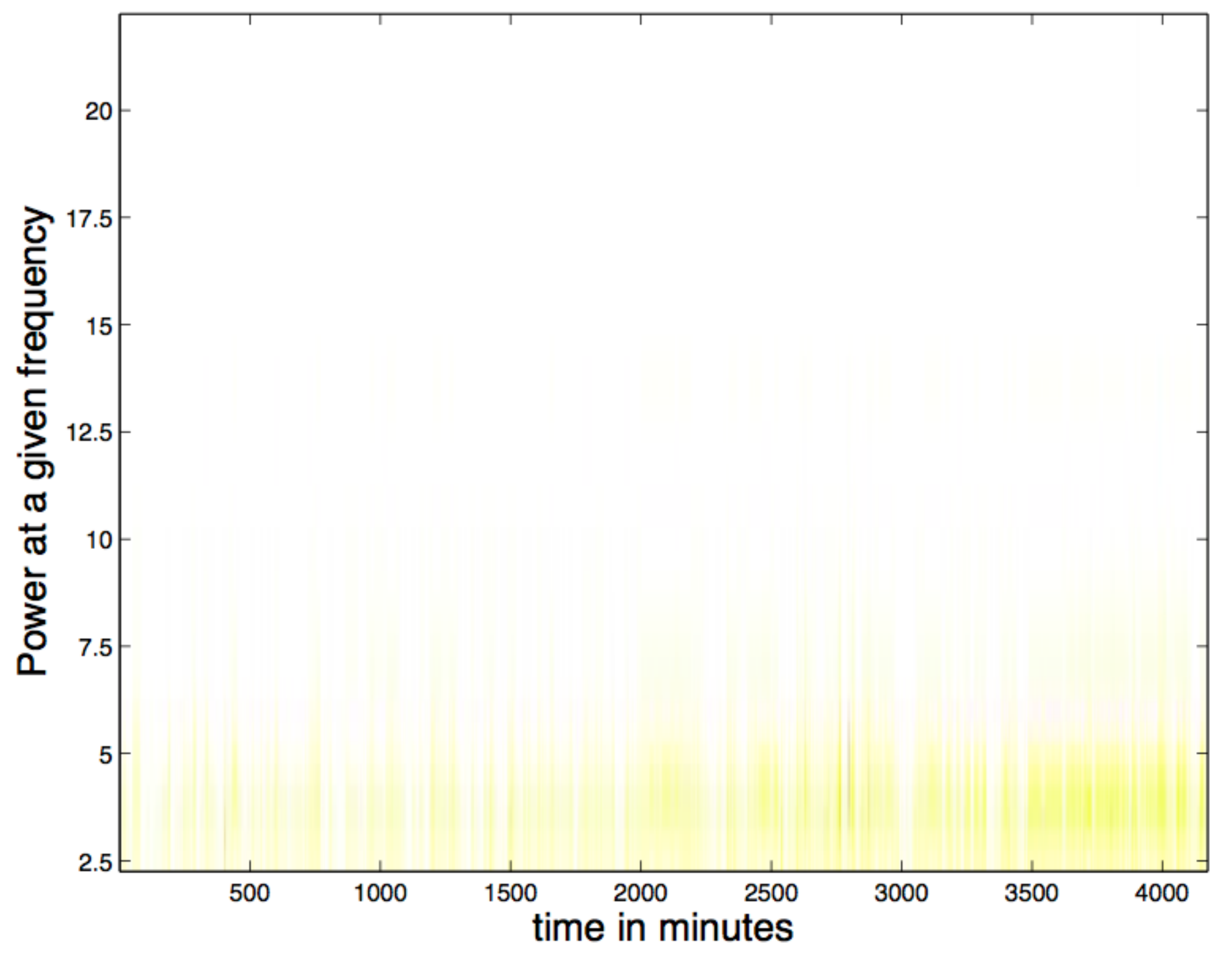}
\includegraphics[width=5cm,height=3cm]{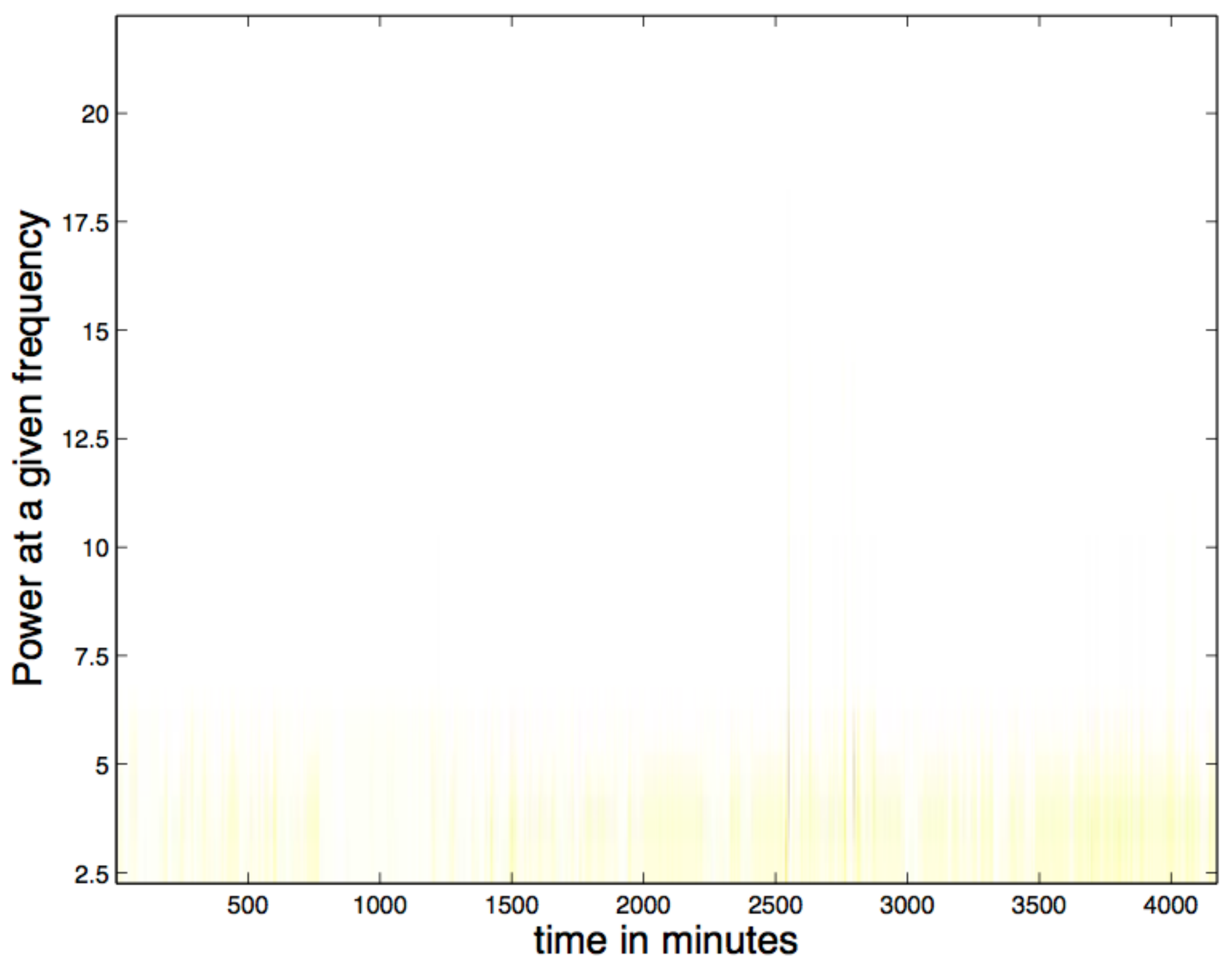}
\includegraphics[width=5cm,height=3cm]{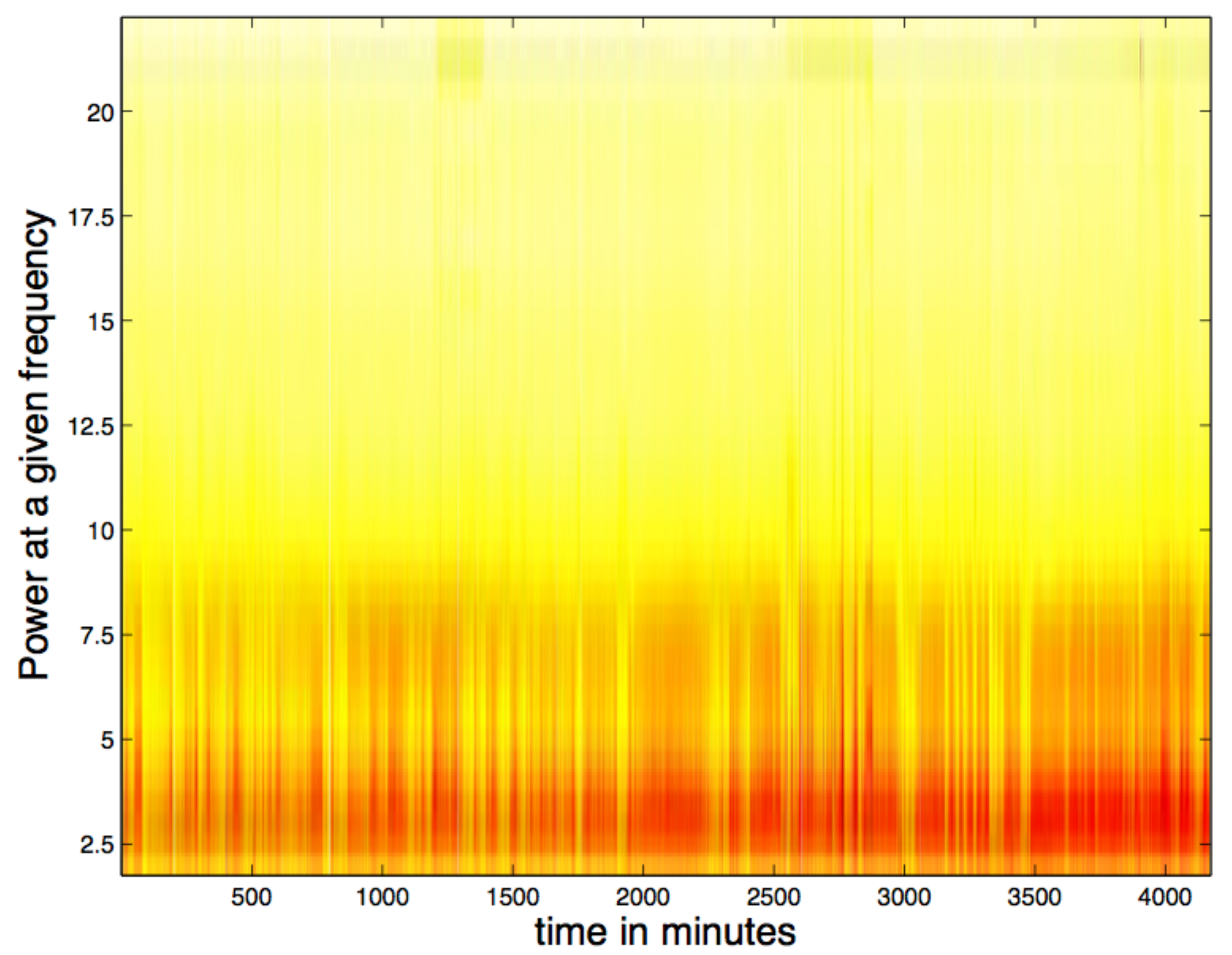}
\caption{\footnotesize  PS of the EEG for the left, right, and whole brain
  (left to right) respectively. Note, the left and right portions figures are not empty, but
  rather the range the variables take is very large and thus difficult
to visualize. Finally, note that no seizure-like event is evident.}
\label{fig:psd_weak}
\end{figure*}

A subset of the SAH patients we study have a depth electrode inserted
in their brain to monitor electrical activity for clinical reasons. A
subset, $18$ of $48$, of these patients displayed a ``depth seizure,''
or a seizure that was not easily identifiable from the surface EEG. In
our example, the depth seizure was manually identified from the raw
EEG collected from the depth probe. Here the PS of the EEG was
recorded once a minute with an $n=40$ spit into half a Hz intervals
ranging from 1-20 Hz for the entire length of the patient stay. Here
we set $T=40$ minutes, making $n=k=40$, allowing again for
$\mathbf{X}(T_k)$ to be of full rank. Because of the lower resolution
of the recordings we ignore fine scale features and work to detect
seizure by identifying abrupt changes in the active
frequencies. Figure \ref{fig:psd_weak} shows the PS of the surface EEG
for the left, right, and whole brain respectively. In these plots, no
discernible seizure-like structure is evident. The EOF of the PS
signals tell another story. In this patient the feature we would hope
to see, abrupt changes in the frequencies with energy, are clearly
visable in the left brain only; neither the whole brain nor the right
brain have a single strong enough to identify a seizure-like event in
the EOF signal. This corroborates what we expect in half of the depth
seizure cases --- a localized seizure event that is not propagating
through the rest of the brain. Therefore, in this example, applying
the EOF to the PS of the surface EEG can help identify depth seizure
events using \emph{surface EEG data} that are only identifiable
manually using the depth EEG and are manually unidentifiable using the
PS of either the depth or the surface EEG alone.

\paragraph{Seizure identification in a small population of SAH
  patients with and without seizure.}

A first step toward using an EOF signal to define a seizure phenotype
is to determine whether an EOF signal can be used to identify seizure
by humans at a broad level.  Using a data set consisting of $26$
patients, $1$ with a surface and $12$ with depth seizures, we compared
EOF visualization to PS visualization in correctly identifying
seizures according to a gold standard generated by a trained
neuroscientist. Accuracy was $88 \%$ for EOF versus $48 \%$ for PS,
and the difference was statistically significantly different by
McNemar's test ($p=0.008$) \cite{mcnemars_test}. Similarly, using the
EOF of the PS there is a strong, statistically significant
\emph{linear} correlation ($\rho = 0.76$, $p=10^{-5}$) between
EOF-identified seizure and the gold standard. There was no linear
correlation between the PS identified seizure and the gold
standard. Relative to the data set here, the EOF was not useful for
differentiating depth versus surface seizure; this lack of
effectiveness is likely due to sample size (correlation for depth
identification was strong but not statistically significant).


\section{Discussion}

EOF analysis of the PS of EEG highlights the characteristics of EEG
that define seizure in SAH patients. Moreover, the EOF of the PS of
the EEG makes manual identification of seizure significantly easier
and more reliable.  As is often the case, multiple levels of data
analysis, e.g., estimating the EOF of the PS of the EEG, can be very
useful for revealing the important temporal content.

Given the likelihood that seizures in SAH patients have different
implications for different populations of patients (e.g., older
patients, patients with great injury, etc.), discovering and
quantifying the differentiating temporal signatures and tying them to
outcomes will be of critical significance for both understanding and
treating seizure in patients with SAH. Nevertheless scientifically
controlled collection of EEG data for SAH patients are rare if
nonexistent. Here we show that it is possible to use physiologic data
collected for clinical reasons to better understand SAH-based
pathophysiology, even though the data are collected outside of a
scientifically controlled environment and contain noise, missing
values, clinical intervention effects, and nonstationary
trends. Nevertheless, much work remains both to automate seizure
detection in SAH patients using the EOF of the PS, and to
statistically define the temporal signatures that can be used to
differentiate patient health and predict patient outcome.


\section*{Acknowledgments}
DJA's and GH's contribution was funded a grant from the National
Library of Medicine, ``Discovering and applying knowledge in clinical
databases'' (R01 LM006910). MS's contribution was funded by by
C.S. Draper Laboratory, Inc. grant number $SC001-0000000642$. We'd
like to thank Professor Sato for discussions and the invitation, and
NOLTA for hosting the conference.

\bibliography{signal_processing_time_series_analysis,neuroscience,aos_refs,analysis,me,statistics}



\end{document}